\documentclass{llncs}
\usepackage[usenames]{color}
\usepackage{epsf}
\usepackage[round]{natbib}
\usepackage{subfigure,floatflt,wrapfig}
\usepackage{graphicx}
\graphicspath{
{figures/}
}

\begin{document}
\title{Scale Invariance of Immune System Response Rates and Times: Perspectives on Immune System Architecture and Implications for Artificial Immune Systems}
\author{Soumya Banerjee\inst{1} \and Melanie Moses\inst{1}}
\institute{Department of Computer Science, University of New Mexico, USA \email{\{soumya,melaniem\}@cs.unm.edu}}

\maketitle

\begin{abstract}
Most biological rates and times decrease systematically with organism body size. We use an ordinary differential equation (ODE) model of West Nile Virus in birds to show that pathogen replication rates decline with host body size, but natural immune system (NIS) response rates do not change systematically with body size. This is surprising since the NIS has to search for small quantities of pathogens through larger physical spaces in larger organisms, and also respond by producing larger absolute quantities of antibody in larger organisms. We call this \textit{scale-invariant detection and response}. We hypothesize that the NIS has evolved an architecture to efficiently neutralize pathogens. We investigate a range of architectures using an Agent Based Model (ABM). We find that a sub-modular NIS architecture, in which lymph node number and size both increase sublinearly with body size, efficiently balances the tradeoff between local pathogen detection and global response using antibodies. This leads to nearly scale-invariant detection and response, consistent with experimental data. Similar to the NIS, physical space and resources are also important constraints on Artificial Immune Systems (AIS), especially distributed systems applications used to connect low-powered sensors using short-range wireless communication. We show that AIS problems, like distributed robot control, will also require a sub-modular architecture to efficiently balance the tradeoff between local search for a solution and global response or proliferation of the solution between different components. This research has wide applicability in other distributed systems AIS applications.
\end{abstract}

\section{Introduction}
\label{intro}
Many emerging pathogens infect multiple host species \citep{woolhouse}, and multi-host pathogens may have very different dynamics in different host species \citep{cable}. Understanding how quickly pathogens replicate and how quickly the natural immune system (NIS) responds is important for predicting the epidemic spread of emerging pathogens. We show that pathogen replication rates decline systematically with host body size, but NIS response times do not increase significantly. We discuss how the decentralized architecture of the immune system facilitates parallel search, enabling NIS response times that do not increase substantially with body size.

The NIS solves a search problem in both physical space and antigen space. The length of the search is determined by the time it takes for a cognate B cell to encounter antigen. This is a difficult search problem since very rare antigen specific NIS cells have to search for small quantities of antigen throughout the body. For example, a mosquito injects $10^{5}$ live virus particles into a vertebrate host that has billions or trillions of cells \citep{styer}. The research we describe here suggests that the time for the immune system to detect and neutralize the pathogen is nearly independent of the size of the organism. We call this \textit{scale-invariant detection and response}. 

This is counter-intuitive, since if we inject a sparrow and a horse with the same amount of pathogen, the immune system of the horse has to search a much larger physical space to find the pathogen, compared to the sparrow. This research models how different potential architectures of the lymphatic network enable the NIS to mount an effective immune response that neutralizes pathogens in time that is independent of host body size. Physical space and resource are also important constraints on Artificial Immune Systems (AIS), especially distributed systems applications used to connect low-powered sensors using short-range wireless communication. Our research shows how the optimal design of such AIS can be informed by architectural strategies employed by the natural immune system.

In addition to the immune system having to search larger spaces in larger organisms, larger body size may be expected to slow immune system response times because the metabolic rate of cells is lower in larger species \citep{brown_mte}. The metabolic rate of each cell is constrained by the rate at which nutrients and oxygen are supplied by the cardio-vascular network. The rate at which this network supplies nutrients to each cell ($R_{cell}$), scales as the body mass ($M$) raised to an exponent of -1/4: $R_{cell} \propto M^{-1/4}$, such that individual cellular metabolic rates decrease as the body mass increases \citep{west_mst,brown_mte}. The metabolic rate of a cell dictates the pace of many biological processes \citep{brown_mte}. This could affect NIS search times by reducing movement and proliferation of immune cells \citep{perelson_scaling}. Rates of DNA and protein synthesis are also dependent on the cellular metabolic rate and could influence the rate at which pathogen replicates inside infected cells \citep{cable}. 

The possibilities that NIS cells and pathogens may move and proliferate at speeds independent of mass ($ \propto M^{0}$) or proportional to cellular metabolic rate ($ \propto M^{-1/4}$) lead to four hypotheses, shown in Table \ref{tab:Tab1}, as originally proposed by Wiegel and Perelson \citeyearpar{perelson_scaling}.

We combine an ordinary differential equation model, an Agent Based Model (ABM) and empirical results from experimental infection studies on West Nile Virus (WNV) \citep{komar,horses_bunning,catsdogs} in what is the first test that we are aware of, examining the effects of body size on pathogen replication and immune system response rates. Our results are consistent with H2: pathogen replication rate $ \propto M^{-1/4}$ and NIS rates $ \propto M^{0}$. 

The remainder of the paper is organized as follows: Section \ref{brief_intro} gives an introduction to the relevant immunology; our statistical methods are outlined in Section \ref{stat}; Section \ref{ode_model} discusses an ordinary differential equation model of pathogen growth and immune response; Section \ref{needle_haystack} discusses the difficulties faced by the NIS in searching space; we use an ABM to derive scaling relations for NIS cell detection and migration times in Section \ref{abm}; the results are summarized in Section \ref{results}; Section \ref{submodular} explains how a sub-modular NIS architecture balances fast search times and fast communication to recruit NIS cells, leading to scale invariant detection; Section \ref{ais} discusses the applications and implications for this work in distributed systems AIS, using an example of multi-robot control, and lastly we make concluding remarks in Section \ref{conclusion}.

\begin{table}
\caption{Four scaling hypotheses of pathogen replication and immune system response rate \citep{perelson_scaling}}
\label{tab:Tab1}
\begin{tabular}{lll}
\hline\noalign{\smallskip}
\noalign{\smallskip}\hline\noalign{\smallskip}
H1: Pathogen replication rate $\propto M^{0}$ & H2: Pathogen replication rate $\propto M^{-1/4}$ \\ 
\hspace{5 mm} NIS search time $\propto M^{0}$ & NIS search time $\propto M^{0}$ \\ \hline
H3: Pathogen replication rate $\propto M^{0}$ & H4: Pathogen replication rate $\propto M^{-1/4}$ \\
\hspace{5 mm} NIS search time $\propto M^{-1/4}$ & NIS search time $\propto M^{-1/4}$ \\
\noalign{\smallskip}\hline
\end{tabular}
\end{table}

\section{A Brief Introduction to the Immune System}
\label{brief_intro}

The NIS has two main components: the innate immune system and the adaptive immune system. The innate immune system is the first line of defense of an organism and consists of complement proteins, macrophages and dendritic cells (DC) \citep{janeway}. The adaptive immune system consists of T helper cells, B cells and cytotoxic T cells. The area of tissue that drains into a lymph node (LN) is called its draining region (DR). The lymphatic system collects extra-cellular fluid called lymph from tissues and returns it to blood \citep{janeway}. DCs sample the tissue in DRs for pathogens, and upon encountering them, migrate to the nearest LN T cell area to present antigen to T helper cells. Cognate T helper cells specific to a particular pathogen are very rare ($1$ in $10^{6}$ NIS cells) \citep{soderberg}. Upon recognizing cognate antigen on DCs, T helper cells proliferate and build up a clonal population in a process called clonal expansion. While proliferating, T helper cells also migrate to the LN B cell area to activate B cells. Cognate B cells specific to a particular pathogen are also very rare. They need to recognize cognate antigen on follicular dendritic cells (FDC) and also need stimulation from cognate T helper cells. After recognition, cognate B cells undergo clonal expansion and differentiate into antibody-secreting plasma cells \citep{janeway}.

This difficult search through the large physical space is facilitated by infected site inflammation, chemokines and preferential expression of adhesion molecules, which guide NIS cells to sites of pathogen invasion \citep{janeway}. The infected site LN recruits NIS cells from other LNs and we refer to the recruitment time as communication overhead between LNs. Some pathogens do not invoke all the arms of the immune system, e.g. some bacteria are efficiently eliminated by the innate immune system. We focus here on pathogens, like WNV, that elicit an antibody response. However, the arguments put forward in this paper would apply also to pathogens that evoke a cytotoxic T cell or other response. 

We are interested in the physical structure of the NIS, and we hypothesize that evolutionary pressures have shaped NIS architecture to minimize the time taken to unite rare antigen-specific NIS cells with their pathogens. This requires both rapid detection of the initial pathogens and also rapid clonal expansion to produce sufficient T helper cells to activate a critical number of B cells. These B cells will then undergo clonal expansion and differentiate into antibody-secreting plasma cells. Hence this paper focuses primarily on the uptake of antigen by DCs in DR, recognition of antigen on DCs by T cells in LN, the subsequent process of clonal expansion and recruitment of B cells from other LNs.

\section{Statistical Methods}
\label{stat}

We use ordinary least squares (OLS) regression to test whether our model predictions are consistent with hypothesized scaling relationships and, where possible, biological measurements. We calculate the $r^{2}$ value, where $r$ is the Pearson correlation coefficient and the $r^{2}$ quantifies the proportion of variation that the independent variable explains in the dependent variable. We test how empirical data from literature and results of our simulation scale with mass by taking the logarithm of both variables and doing an OLS regression. We report whether the scaling exponent is consistent with -0.25, 0, or both. We test for significance at the alpha = 0.05 level. The mean is reported after testing all log-transformed datasets for normality using the Jarque-Bera test \citep{jarque_bera}.

\section{An Ordinary Differential Equation Model for Viral Dynamics}
\label{ode_model}

A standard Ordinary Differential Equation (ODE) model was developed to observe how viral proliferation rates and immune response rates scale with body size, and model results were parameterized to empirical levels of virus in blood \citep{icarishybrid}. Data on viral proliferation was taken from studies, which used the same West Nile Virus (WNV) strain to experimentally infect 25 different species with body mass ranging from 0.02 kg (house finch) to 390 kg (horse), and the viral load was monitored each day in blood serum over a span of 7 days post infection (d.p.i) \citep{komar,horses_bunning,catsdogs}. In the model, $p$ = rate of virion production per infected cell, $\gamma$ = innate immune system mediated virion clearance rate, $\omega$ = adaptive immune system proliferation rate, and $t_{pv}$ = time to attain peak viral concentration. Equations (1) to (4) are shown below

\begin{equation}
\frac{dT}{dt} = -\beta TV
\end{equation}
\begin{equation}
\frac{dI}{dt} = \beta TV - \delta I
\end{equation}
\begin{equation}
\frac{dV}{dt} = pI - c(t)V
\end{equation}
\begin{equation}
c(t) = \left\{ \begin{array}{rcl}
	\gamma & \mbox{,}  & t < t_{pv} \\
	\gamma e^{\omega (t - t_{pv})} & \mbox{,} & t \geq t_{pv}
	\end{array}\right.
\end{equation}

Target cells $T$ are infected at a rate proportional to the product of their population size and the population size of virions $V$, with a constant of proportionality $\beta$. The loss in the target cell population is balanced by a gain in the infected cell population. Infected cells $I$ also die at a rate $\delta I$. Virions are produced by infected cells at a rate $p$ and cleared by the immune system at the rate $c(t)V$. The action of the immune system is decomposed into an innate response ($\gamma$) before the virus concentration attains a peak, and an exponential adaptive immune response after peak due to clonal expansion characterized by a proliferation rate $\omega$. This paper focuses on $p$ as pathogen replication, and $\gamma$ and $\omega$ as immune system response.

Empirical data show that time to peak viral concentration ($t_{pv}$, time between infection and peak viral concentration in blood) for WNV empirically occurs between 2 - 4 d.p.i. If this peak were due to target cell limitation, then we would expect $t_{pv}$ to increase with host mass $M$ since larger animals have more target cells. However, $t_{pv}$ is likely to be determined by WNV specific antibodies, which have a critical role in WNV clearance \citep{diamond_igm}. If the peak is determined by a threshold presence of antibodies, then it implies that $t_{pv}$ is determined by the time for cognate B cells to recognize antigen, proliferate and produce antibodies: $t_{pv} = t_{detect} + t_{prolif}$. Empirically, $t_{pv}$ is independent of host mass \citep{komar}. The time $t_{pv}$  is highly conserved, ranging only between 2  and 4 d.p.i. across different hosts that range in mass from 0.02 kg (house finch) to 390 kg (horses). Since there is no statistically significant relationship between $t_{pv}$ and $M$ (p-value testing significance of slope not equal to zero - 0.35, 95\% CI on slope - [-0.037, 0.0607]), then the data are consistent with the hypothesis that $t_{pv} \propto M^{0}$. 

Hence empirical data for WNV supports the hypothesis that NIS response rates are independent of $M$. Other empirical data suggests that pathogen replication rates scale as $M^{-1/4}$ for a variety of pathogens in a variety of hosts, including WNV in birds and mammals \citep{cable}. Together, these observations reject all hypotheses in Table \ref{tab:Tab1}, except H2: pathogen replication declines with $M^{-1/4}$ and immune response times are invariant with respect to $M$. Ideally we would test both hypotheses simultaneously using the ODE model described by Equations (1) to (4). However, the ODE has too many parameters to simultaneously test both hypotheses without overfitting the data. Since our main goal is to determine the scaling of the NIS response rates and times, we fit the data by choosing initial parameter estimates consistent with pathogen replication rates scaling as $M^{-1/4}$, and then fit the model to the data.

\begin{figure}[ht!]
 \includegraphics[width=1\textwidth]{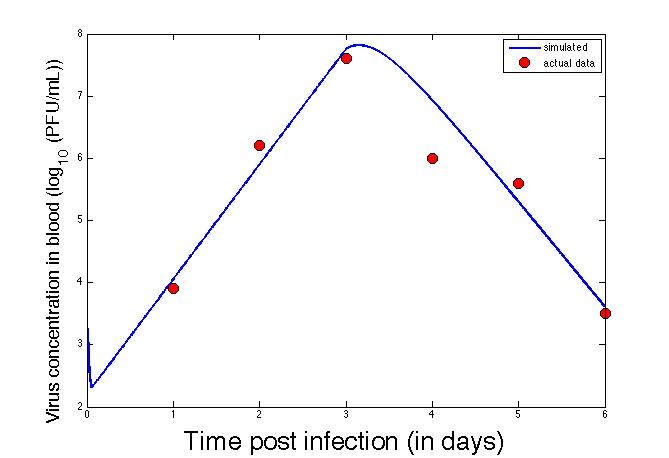}
\caption{A sample plot of virus concentration in blood vs. time post infection (solid line - predicted ODE model output, circles - actual experimental data for a great-horned owl \citetext{data from \citealp{komar}}. Y axis - virus concentration in $log_{10}$ PFU/mL of blood, X axis - days post infection.} 
 \label{fig:Fig1}
\end{figure}

The ODE model was fit to the viral load data for each of 25 species \citep{komar,horses_bunning,catsdogs} for days 1 - 7 d.p.i and the model parameters were estimated using non-linear least squares regression. The Berkeley Madonna\textregistered \citep{madonna} software package was used to generate the fits and we assigned the parameter value to the mass of the species to do an OLS regression. The scaling relations found were:
$p \propto M^{-0.29}$ (the predicted exponent of -0.25 is in the 95\% CI, $r^{2}$ = 0.31, p-value testing significance of slope = 0.0038). The innate immune system mediated pathogen clearance rate ($\gamma$, $day^{-1}$) and adaptive immune system cell proliferation rate ($\omega$, $day^{-1}$) were independent of host mass $M$ (p-values testing significance of slope not equal to zero - 0.4238 and 0.7242 respectively, 95\% CI on slope - [-0.04, 0.146] and [-0.347,0.4319] respectively). In ongoing work, we are using simpler models with fewer parameters to simultaneously fit replication and NIS response and are getting similar predictions without seeding the initial estimates of $p$ to be proportional to $M^{-1/4}$. A sample parameter estimate and model prediction is shown in Table \ref{tab:Tab2} and Fig. \ref{fig:Fig1}.

\begin{table}
\renewcommand{\arraystretch}{1.8}
\setlength{\tabcolsep}{4pt}
\caption{Estimated ODE model parameters for great-horned owl (PFU - Plaque Forming Units}
\label{tab:Tab2}
\begin{tabular*}{1\textwidth}{|ccccccc|}\hline
$V_{0}\left(\frac{\mbox{PFU}}{\mbox{mL}}\right)$ & $\beta \left(\frac{1}{\frac{\mbox{PFU}}{\mbox{mL}}\mbox{day}} \right)$ 
& $p \left(\frac{1}{\mbox{PFU}\mbox{day}}\right)$ & $\delta \left(\frac{1}{\mbox{day}}\right)$ & $\gamma\left(\frac{1}{\mbox{day}}\right)$ 
& $\omega\left(\frac{1}{\mbox{day}}\right)$ & $t_{pv}\left(\frac{1}{\mbox{day}}\right)$\\\hline
2.95 & $10^{-7}$ & 524.97 & 1.19 & 91.99 & 2.7 & 3\\\hline
\end{tabular*}
\end{table}

These findings from our ODE model and empirical data are consistent with hypothesis H2: pathogen replication rates decline in larger hosts, but immune response is independent of host mass (scale-invariant detection and response). This raises the question: what mechanisms make NIS rates independent of host body mass and metabolism? The problem of slower metabolism in larger hosts can be circumvented in immune response if LNs have \textit{privileged metabolism} i.e. they consume energy which is independent of host mass \citep{perelson_scaling}. However, even if immune system cells are not constrained by the lower mass specific metabolism in larger organisms, it remains to be explained how larger spaces can be searched in invariant time.

\section{Searching for a Needle in a Haystack}
\label{needle_haystack}

The NIS is confronted with a very difficult search problem. Extremely rare B cells or T cells specific to a particular antigen ($1$ antigen-specific T cell in $10^{6}$ T cells)\citep{miller} search for initially rare antigen in localized tissue. A constant number of virions is injected into a host by mosquitoes, regardless of host size, for WNV \citep{styer} and other pathogens spread by mosquito vectors. Our analysis of the ODE model and the empirical $t_{pv}$ suggest that the NIS in bigger organisms can find this fixed number of virus particles in approximately the same time as in smaller organisms i.e. a horse finds those virions in a haystack 10,000 times larger than a sparrow's haystack, but in the same time.

By way of introduction we define a \textit{completely modular system} as one that is composed of self-contained units that are a fixed size and do not need to communicate, and a \textit{module} as a LN and its DR. To simplify our models, we assume that each LN has a single DR and a DR drains into a single LN. Perfectly parallel search is easily achievable if immune response is completely modular at the LN level and systemic communication in the immune system does not generate more overhead in larger systems. Completely modular systems have no overhead of communication and hence achieve perfectly parallel search \citep{amdahl} since search is in a space of the same size and is replicated in parallel. However, experimental evidence suggests that the NIS is not modular at the LN level (see Section \ref{submodular}) \citep{halin,altman_dittmer,hildebrandt}.

A conceptually similar example of modularity can be found in the concept of a \textit{protecton} \citep{protecton}, which is a modular unit of protection consisting of $10^{7}$ B cells of different specificities per mL of volume and is iterated proportional to the size of the organism i.e. if one samples 1 mL of a tadpole and 1 mL of an elephant, we will likely find the same set of $10^{7}$ B cells but the elephant will have more copies of the protecton. This modular design reduces the time taken to build up a population of effector cells by clonal expansion. However, we would like to point out that the concept of a protecton is theoretical. The theory states that the NIS could be \textit{constructed} in a modular fashion. The NIS cells constituting a protecton however can follow different migration paths and communicate with cells outside their group. Nevertheless, this further motivates the question of investigating whether there is modularity at the level of LN that would help to parallelize the search process. Sections \ref{abm} and \ref{results} explore the empirical architecture of LN organization, and explain why a purely modular architecture is not optimal.

\section{Agent Based Model to Explore How LN Size Affects NIS Response Time}
\label{abm}

The general model of immune system dynamics in the LN and its DR are shown in Fig. \ref{fig:Fig2} and summarized as follows:

\begin{figure}[ht!]
 \includegraphics[width=1\textwidth]{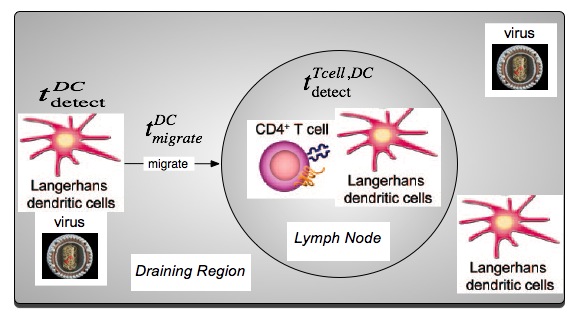}
\caption{Immune system dynamics within a lymph node and its draining region.} 
 \label{fig:Fig2}
\end{figure}

\begin{enumerate} 
\item Stage 1: DCs randomly search for antigen in a local DR. The time taken to detect antigen is denoted by $t^{DC}_{detect}$.
\item Stage 2: DC migrates to a local LN T cell area along a chemotactic gradient. The time taken to migrate is $t^{DC}_{migrate}$.
\item Stage 3: Antigen-specific T cell in a LN detects antigen on DC and the time taken to detect is $t^{Tcell,DC}_{detect}$. T cells then activate cognate B cells, which undergo clonal expansion, to produce antibody-secreting plasma cells.
\end{enumerate}

Total time to detect antigen is given by
\begin{equation}
t_{detect} = t^{DC}_{detect} + t^{DC}_{migrate} + t^{Tcell,DC}_{detect}
\end{equation}

The total time spent in communication (recruiting other NIS cells to the LN) by the draining infected site LN is explained in detail in the Section \ref{submodular} and is shown below

\begin{equation}
t_{comm} = N_{comm}/rate_{comm} \propto M/V^{2}_{LN}
\end{equation}

We use an ABM to investigate how organization of the lymphatic network minimizes the time to detect antigen $t_{detect}$ and $t_{comm}$, leading to scale-invariant detection and response.

We note that the total volume of organs and fluids in mammals both scale proportional to $M$ \citep{peters,west_mst}, so in our models, we assume that LN volume and the total volume of lymph in the entire body are both proportional to $M$. Thus, the number of LN multiplied by the volume of each LN is proportional to $M$. Additionally, the volume of a DR is determined by $M$ (which is proportional to body volume, given that tissue density is constant across animals)\citep{peters} divided by the number of LN, i.e., for a fixed size $M$, more LN result in a smaller DR for each LN. 

In order to explore how the spatial arrangement of LNs affects time to detect antigen, we use the CyCells \citep{cycells,cycells_url} ABM to simulate viral replication in a 3D compartment representing the LN and DR. We simulated DCs, T cells, viruses and LNs, and explicitly modeled DC migration from tissue to LN along a chemotactic gradient, and random walk of DC and T cells in LN T cell area. The model parameters are summarized in Table \ref{tab:Tab3}.

The assumptions that we make are: a) initially, we ignore migration of antigen specific B cells from other LNs. We consider how such systemic responses change the NIS architecture in the next section, b) there is a fixed chemotactic gradient for DCs to migrate into the LN, c) the pathogen does not replicate until it is in the LN as is the case for WNV, d) DCs and T cells perform random walks in LN T cell area \citep{bajenoff}, and e) LNs have \textit{preferential metabolism} \citep{perelson_scaling}, i.e. inside a LN, NIS cells have speed and proliferation rates that are invariant with host mass $M$.

{\tiny
\begin{table}
\renewcommand{\arraystretch}{1.5}
\setlength{\tabcolsep}{5pt}
\caption{The parameters used in the agent based model together with a short description of their role and default value (L - literature, F - fit to data)}
\label{tab:Tab3}
\begin{tabular*}{1\textwidth}{| p{4.5cm} c  p{4cm} c p{3.5cm} c |}\hline
Description & Value & Source  \\ \hline
Side length of cubic compartment (DR) & $1000\mu m$ to $4000 \mu m$ & L \citep{halin}\\
Side length of cubic compartment (LN) & $250\mu m$ to $1000 \mu m$ & L \citep{halin}\\
Duration of a time step & 60 $sec$ & F\\
Number of antigen specific B cells in a lymph node of $10^{6}$ B-cells & 1 & L \citep{miller}\\
Density of DC in DR & $1250/mm^{3}$ & L \citep{banchereau}\\
Amount of antigen in DR & 100 & L \citep{styer}\\
Radius of T cell & $10\mu m$ & L \citep{miller}\\
Radius of antigen-presenting DC & $30\mu m$ & L \citep{miller}\\
Speed of T cell & $0.1664\mu m/sec$ & L \citep{miller}\\
Speed of antigen-presenting DC in LN & $0.0416\mu m/sec$ & L \citep{miller}\\
Speed of antigen-presenting DC in DR & $0.0832\mu m/sec$ & L \citep{jolanda}\\
Sweep and sense distance of antigen-presenting DC (measured from cell center) & $50\mu m$ & L \citep{miller}\\
\hline
\end{tabular*}
\end{table}
}

\section{Results}
\label{results}

We first use an ABM to calculate the detection and migration times in Eqn. (6) for mice and then show how each of these times scale with LN and DR dimensions. We then use these scaling relations to derive analytical expressions for detection, migration and communication times for three hypotheses of LN organization.

\subsection{The Base Case Model of a Typical Lymph Node and Scaling Up}
\label{base_case}

A DR was simulated as a cubic compartment of side $4000\mu m$ with a cubic LN of length $1000\mu m$. The mean time taken by DCs to detect antigen and the time taken by antigen-specific T cell to recognize antigen on DC are shown in Table \ref{tab:Table4}. These are in agreement with experimental observations in mice \citep{miller}. The total time to detect antigen ($t_{detect} = t^{DC}_{detect} + t^{DC}_{migrate} + t^{Tcell,DC}_{detect}$) is then around 19 hours which is in agreement with experimental studies in mice \citep{itano} and consistent with our observation that peak viral concentration occurs in day 2 to 4 for WNV across organisms- since we hypothesize that the peak occurs due to WNV-specific NIS cells, which must first have come into contact with WNV in the LN.

The DR and LN regions were then scaled up and we observed how DC detection, migration and T cell interaction times scaled with the size of the DR and LN. We simulated 3 cubic DRs: DR of length $1000\mu m$ with LN of length $250\mu m$, DR of length $2000\mu m$ with LN of length $500\mu m$, and DR of length $4000\mu m$ with a LN of length $1000\mu m$. The observed scaling relations are in Table \ref{tab:Table5} and are consistent with DC migration time scaling with the mean distance from DR to LN. The time for DC to detect antigen specific T cell in LN was found to be uncorrelated with the size of the LN. 

\begin{figure}[ht!]
\includegraphics[height=80mm]{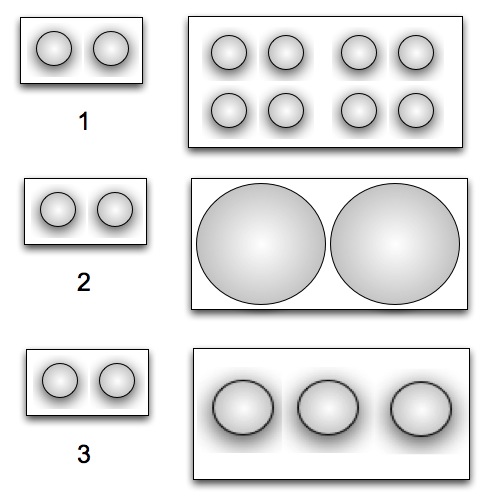}
\caption{The 3 different hypotheses of scaling of lymph node size and numbers. (1) Completely Modular Detection Network: base organism with 2 LNs and another organism 4 times as big with 4 times the number of lymph nodes each of the same size as the base organism. (2) Non-Modular Detection Network: organism 4 times bigger has the same number of LNs but each is 4 times bigger. (3) Hybrid Sub-Modular Detection Network: organism 4 times bigger has more LNs each of which are also bigger.} 
 \label{fig:Fig3}
\end{figure}

We now explore 3 competing hypotheses of lymphatic system organization (Fig. \ref{fig:Fig3}). 

{\tiny
\begin{table}
\renewcommand{\arraystretch}{1.5}
\setlength{\tabcolsep}{10pt}
\caption{Simulated values in mice for DC antigen detection, DC migration and DC-T cell interaction}
\label{tab:Table4}
\begin{tabular*}{1\textwidth}{| p{3cm} c  p{4cm} c  c |}\hline
Times & Dimensions (DR, LN) & Value (simulation)\\ \hline
$t^{DC}_{detect} + t^{DC}_{migrate}$ & $4000\mu m$, $1000 \mu m$ & Mean = 4.13 hours, SD = 1.4 hours, 10 simulations\\
$t^{Tcell,DC}_{detect}$ & $4000\mu m$, $1000 \mu m$ & Mean = 15.13 hours, SD = 6 hours, 10 simulations\\
\hline
\end{tabular*}
\end{table}
}

{\tiny
\begin{table}
\renewcommand{\arraystretch}{1.5}
\setlength{\tabcolsep}{5pt}
\caption{Scaling relations for time for DC antigen detection, DC migration and DC-T cell interaction ($r_{DR}$ = radius of DR, $r_{LN}$ = radius of LN). \(^+\) - testing for significance of exponent = 1, \(^{\S}\) - testing for significance of exponent $\neq$ 0. }
\label{tab:Table5}
\begin{tabular*}{1\textwidth}{| p{2.5cm} c   c p{4cm} c p{2.5cm} c |}\hline
Times & Dimensions (DR) & Scaling Relation & Statistics \\ \hline
$t^{DC}_{detect} + t^{DC}_{migrate}$ & $1000\mu m$, $2000 \mu m$, $4000 \mu m$ & $\propto (r_{DR} - r_{LN})^{0.91}$ & $r^{2} = 0.98, p < 0.001 $ \(^+\)\\
$t^{Tcell,DC}_{detect}$ & $1000\mu m$, $2000 \mu m$, $4000 \mu m$ & $\propto r^{0}_{LN}$ & $p > 0.05$ \(^{\S}\) \\
\hline
\end{tabular*}
\end{table}
}

{\tiny
\begin{table}
\renewcommand{\arraystretch}{1.5}
\setlength{\tabcolsep}{5pt}
\caption{Scaling relations for LN and DR parameters and how $t^{DC}_{migrate}$ depends on DR and LN dimensions ($N$ - number of LNs, $V_{lymph}$ - volume of lymph, $V_{LN}$ - volume of LN). \(^{\S}\) - taken from empirical data \citep{altman_dittmer}}
\label{tab:Table6}
\begin{tabular*}{1\textwidth}{| p{1cm} c p{2.5cm} c p{1cm} c p{1cm} c p{2cm} c p{1cm} c p{1cm} c |}\hline
LN Architecture & $N$ & $V_{lymph} \propto N*V_{LN}$ & $V_{LN}$ & $V_{DR} \propto M/N$ & $t^{DC}_{migrate} \propto r_{DR} - r_{LN}$ & $t_{comm} \propto M/V^{2}_{LN}$\\ \hline
Model 1 & $M^{1}$ & $M^{1}$ & $M^{0}$ & $M^{0}$ & $M^{0}$ & $M^{1}$\\
Model 2 & $M^{0}$ & $M^{1}$ & $M^{1}$ & $M^{1}$ & $M^{1/3}$ & $M^{-2}$\\
Model 3 & $M^{1/2}$\(^{\S}\) & $M^{1}$ & $M^{1/2}$ & $M^{1/2}$ & $M^{1/6}$ & $M^{-1/2}$\\ \hline
\end{tabular*}
\end{table}
}

\subsection{Model 1: Completely Modular Detection Network}
\label{model1}

In our first model, we assume that the lymphatic network forms a \textit{completely modular network} containing LN of constant size and number of LN proportional to organism size $M$. Using the scaling relationship for $t^{DC}_{migrate}$ from Table \ref{tab:Table5} and noting that in this model the LN and DR dimensions do not change with organism size gives us  $t^{DC}_{migrate} \propto M^{0}$. Since the ABM predicts that detection times in LN do not depend on LN dimensions (Table \ref{tab:Table5}) we have $t_{detect} \propto M^{0}$ and hence the completely modular architecture gives us perfect scale-invariant detection. The predicted relations are summarized in Table \ref{tab:Table6}.

\subsection{Model 2: Non-Modular Detection Network}
\label{model2}

The second model is the other extreme, that LNs are arranged in a detection network with a constant number of LN (all animals have the same number of LNs, however the size of LNs is larger in larger animals). In this model, the DR volume increases proportional to organism mass, and the average distance that a DC has to travel from the DR to the LN increases with organism mass as $M^{1/3}$ (Table \ref{tab:Table6}). Hence $t^{DC}_{migrate} \propto M^{1/3}$ and since DC migration times are around 4 hours in mice, this model predicts that DC migration times in horses (which are 25,000 larger than mice) will be 30 times more than that in mice, which is 5 days in horses and greater than the total observed time for antibody response. Hence the hypothesis of complete lack of modularity in the NIS is ruled out.

\subsection{Model 3:  Hybrid Sub-Modular Architecture}
\label{model3}

This architecture lies midway between Model 1 and Model 2. In this model LNs increase in both size and in numbers as animal size increases, and so does the size of the DR. Table \ref{tab:Table6} gives the predicted relation for $t^{DC}_{migrate} \propto M^{1/6}$. This will lead to migration times that is only 5 times longer in horses (around 20 hours) than in mice. It is not implausible that detection should take so long in a horse, and these slight increases in detection time might be compensated for by slower rates of exponential growth by the pathogen, as predicted by our ODE model (Section \ref{ode_model}). Hence the sub-modular architecture produces detection times which are consistent with our empirical observations (scale-invariant detection), since the difference between 4 hours and 20 hours cannot be resolved on the basis of measurements of viral load taken every 24 hours.

\section{Sub-Modular Architecture Balances Tradeoff Between Local and Global Communication}
\label{submodular}

The few published empirical data that we could find suggest that the mammalian NIS has a hybrid sub-modular architecture (Model 3). There is a trend of increasing LN size and number as animal size increases, for example, 20g mice have 24 LN averaging 0.004g each, and humans are 3000 times bigger and have 20 times more LN, each 200 times bigger \citep{halin,altman_dittmer}. Data from elephants (with LN approaching the size of an entire mouse) and horses (with 8000 LN) also support the hypothesis that LN size and number both increase with body size (Model 3) \citep{hildebrandt,altman_dittmer}; however, data for more species are required to statistically reject any of our 3 models.

We hypothesize that the NIS is submodular (consistent with Model 3) because it is selected not just to minimize time to detect pathogens (achieved by Model 1), but also to minimize the time to produce a sufficient concentration of antibody in the blood ($Ab_{crit}$). A horse 25,000 times larger than a mouse must generate 25,000 times more absolute quantities of antibody ($Ab$) in order to achieve the same concentration of antibody in the blood (where blood volume is $ \propto M$)\citep{peters}. A fixed antibody concentration is required to fight infections like WNV that spread systemically through the blood. We can now consider that the NIS has evolved to minimize two quantities: the time to detect antigen ($t_{detect}$), and the time ($t_{produce}$) to produce $Ab$, where $Ab$ is proportional to $M$ ($Ab$ is the absolute quantity of antibody required to neutralize the pathogen in blood).

In all pathogens which evoke the adaptive immune system, the rate limiting step is the recognition of antigen on DCs by antigen-specific T helper cells within the LN T cell area \citep{soderberg}. The time taken in this recognition step impacts on other downstream processes like activation of B cells since T helper cells activated by DCs must migrate to the B cell area to activate B cells. If organisms of all body sizes activated the same number of initial B cells prior to clonal expansion, the time for a fixed number of B cells to produce $Ab$ would be $\propto log_{2}M$ (since B cells reproduce exponentially through clonal expansion). For example, since it takes 4 days of exponential growth of activated B cells to produce sufficient anti-WNV neutralizing antibody in mice \citep{diamond_igm}, then the corresponding time for a horse would be more than 2 months, should the same number of initial B cells be activated prior to clonal expansion. This conflicts with empirical data on horses \citep{horses_bunning}. We assume that the NIS of larger organisms has to activate a larger number of initial B cells ($B_{crit}$) $\propto M$, in order to build up the critical density of antibodies in a fixed period of time. We now ask how the NIS can activate $B_{crit}$ in our three models.

In Model 1, LNs are a fixed size, and therefore contain a fixed number of B cells, and the smallest LNs (e.g. in mice) contain on the order of a single B cell that recognizes any particular pathogen. Thus, activating $B_{crit}$ to fight an infection like WNV that is initially localized in a single DR, requires recruiting B cells from distant LN. We consider this activation of B cells from remote LN as communication overhead. In general, the number of LNs that a single infected site LN has to communicate with ($N_{comm}$) in order to recruit more B cells is proportional to the amount of antibody required to neutralize the pathogen divided by the number of B cells resident in a LN ($Num_{Bcell}$):

$N_{comm} \propto Ab / Num_{Bcell}$.  Noting that $Ab \propto M$ and $Num_{Bcell} \propto V_{LN}$  we have
$N_{comm} \propto M / V_{LN}$

The rate at which new B cells from other LN enter into infected site LN through the blood or lymphatic vessels ($rate_{comm}$) is proportional to the volume of the LN, $rate_{comm} \propto V_{LN}$ (assuming that a larger LN will have proportionally more high endothelial venules). The time spent in communicating with other LNs and recruiting and activating other B cells ($t_{comm}$) is then given by

\begin{equation}
t_{comm} = N_{comm}/rate_{comm} \propto M/V^{2}_{LN}
\end{equation}

Hence in Model 1, there are increasing costs to communicating with other LN as the organism gets bigger ($t_{comm} \propto M$); it carries out efficient search but is not optimized for antibody production. 

Model 2 (non-modular detection network) compensates for the limitation of physically transporting NIS cells over larger distances by making LNs bigger in larger organisms ($V_{LN} \propto M$). This increases the rate of influx of B cells ($rate_{comm} \propto V_{LN} \propto M$) and also situates more NIS cells inside the infected site LN. Since all the necessary NIS cells that need to be activated are within the LN, this architecture has no communication cost. However, as shown above (Section \ref{model2}), Model 2 leads to DC migration times that are prohibitively long for large animals ($t^{DC}_{migrate} \propto M^{1/3}$). 

The architecture that strikes a balance between the two opposing goals of antigen detection (local communication) and antibody production (global communication) is Model 3 (hybrid sub-modular architecture). It minimizes $T = t_{detect} + t_{produce}$, where $t_{detect}$ = time taken to detect antigen, and $t_{produce}$ = time taken to produce antibody = $t_{comm}$ (since the time taken to produce antibodies is equal to the time taken to recruit B cells or communicate with other LNs; after recruitment starts and cognate T cells recognize antigen on DCs, T cells can migrate to the B cell area and activate B cells in parallel to the recruitment process described earlier). 

We can solve for the total time ($T = t_{detect} + t_{produce}$) to detect antigen and produce B cells using Eq. (7) and the scaling relationship for $t^{DC}_{migrate}$ from Table \ref{tab:Table5} giving:
\begin{equation}
T = t^{DC}_{detect} + a(r_{DR} - r_{LN}) + t^{Tcell,DC}_{detect} + b M / V^{2}_{LN} 
\end{equation}
where $a$ and $b$ are constants, $r_{DR}$ = radius of DR, $r_{LN}$ = radius of LN, $V_{LN}$ = volume of LN, $M$ = organism mass, and $N$ = number of LNs. This becomes 
\begin{equation}
T = t^{DC}_{detect} + cV^{1/3}_{LN} + t^{Tcell,DC}_{detect} + b M / V^{2}_{LN}
\end{equation}

where $c$ is a constant, and $r_{DR}$ and $r_{LN}$ scale as $V^{1/3}_{DR}$ and $V^{1/3}_{LN}$ respectively since $V_{LN}$ and $V_{DR}$ scale isometrically, as $M/N$.

Differentiating with respect to $V_{LN}$ and setting the derivative to zero to find the minimum $T$, gives
\begin{equation}
	V_{LN} \propto M^{3/7}
\end{equation} 
Since the amount of lymph is proportional to host body mass ($N * V_{LN} \propto M$), then 
\begin{equation}
	N \propto M^{4/7}
\end{equation}

These predictions are consistent with the few empirical data we were able to obtain. We note that Wiegel and Perelson predicted $N \propto M^{1/2}$ \citep{perelson_lymphocyte}, but that analysis did not explicitly consider an optimization to simultaneously minimize detection time and time to produce the critical number of $Ab$. 

    In summary, due to the requirement of activating increasing number of NIS cells for antibody production in larger organisms, there are increasing costs to global communication as organisms grow bigger. The semi-modular architecture (Model 3) balances the opposing goals of detecting antigen using local communication and producing antibody using global communication. This leads to optimal antigen detection and antibody production time, and scale-invariant detection and response.

\section{Relevance to Artificial Immune Systems}
\label{ais}

The natural immune system (NIS) utilizes an architecture that functions within constraints imposed by physical space. Physical space is also an important constraint on artificial immune systems (AIS), especially in applications used to connect inexpensive low-powered sensors using short-range wireless communication \citep{kleinberg}. Such spatial networks are being increasingly used in environmental monitoring, disaster relief and military operations \citep{kleinberg}. These networks might operate under constraints of resource and physical space, similar to an NIS. Although there are systematic differences between an NIS and an AIS \citep{timmis_immuno}, the design of the AIS can be informed by architectural strategies employed by their biological counterpart.

\subsection{Original System}
\begin{figure}[ht!]
 \includegraphics[width=1\textwidth]{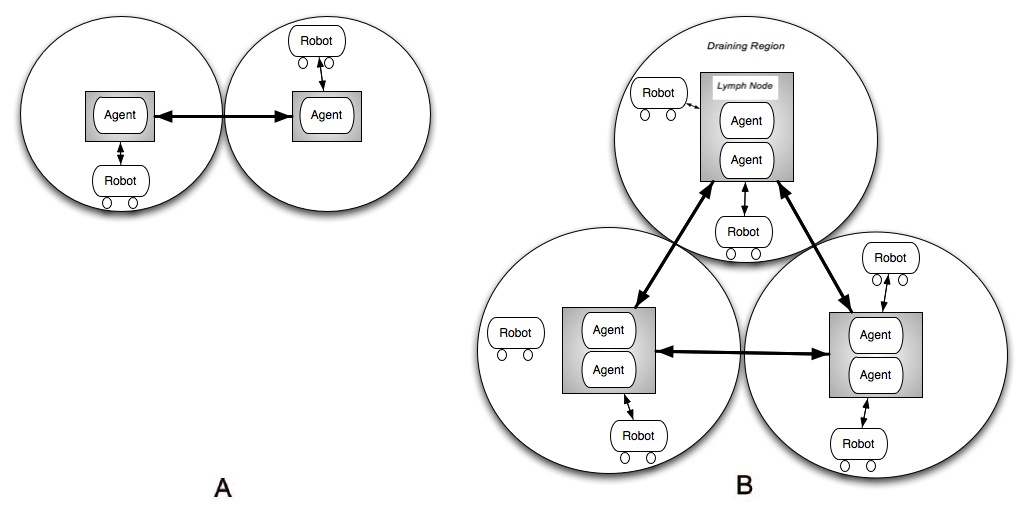}
\caption{(A) Left Panel: a scaled down version of the multi-robot AIS system. The shaded regions are artificial LNs (computer servers) and the unshaded regions are the artificial DR. Light arrows denote communication between robots and servers (local communication) and bold arrows denote communication between servers (global communication). (B) Right Panel: a scaled up multi-robot AIS system with sub-modular architecture. Note that the number of artificial LNs and their size (the number of robots they service and the number of software agents they have in memory) both increase with the size of the system.} 
 \label{fig:Fig4}
\end{figure}

As a concrete example of an application where space is a constraint and scaling of performance with system size is an important design criterion, we consider an AIS approach to control multiple robots tasked with obstacle avoidance \citep{multirobot}. The robots communicate with software agent(s) in a server upon encountering an obstacle. The agents transmit rule-sets of actions to robots to help overcome their obstacles, and agents also share information globally amongst themselves by migrating to other computer servers. Some analogies between this AIS and an NIS are: the obstacle problem presented by a robot is analogous to an antigen, the rule-set of actions transmitted by an agent correspond to antibodies, the robots are akin to DCs, software agents correspond to B cells, the computer servers themselves are analogous to LNs, and the physical area "serviced" by a single computer server corresponds to a DR. The system is diagrammed in Fig. \ref{fig:Fig4} \citetext{modified from \citealp{multirobot}}.

\subsection{Modifying the Original System Using a Sub-Modular Architecture}

We are interested in an architecture that minimizes the time taken by a robot to transmit information about an obstacle (local detection), the time taken by a computer server to transmit back an initial rule-set of actions (local response) and the time taken by a computer server to communicate good rule-sets to other agents (global response). There are two potential communication bottlenecks (communication between robots and computer servers, and communication between computer servers) 

A bottleneck in (local) communication between robot and server demands many small DRs. A bottleneck in (global) server communication requires a few large servers. If both local and global communication are constrained, the architecture which balances these opposing requirements is sub-modular, i.e. the number of servers increases sublinearly with system size and the capacity of each server (bandwidth, memory and number of robots serviced by each server) increases sublinearly with system size (shown in Fig. \ref{fig:Fig4}). The four ways in which communication can be bottlenecked are outlined below:

\begin{enumerate}
\item Unlimited Robot Bandwidth, Unlimited Server Bandwidth: Assuming that robots have unlimited bandwidth to communicate with computer servers and software agents can communicate with each other over a channel with unlimited bandwidth, we see that trivially any of the architectures would suffice.
\item Limited Robot Bandwidth, Unlimited Server Bandwidth: Assuming communication between robots and computer servers is a bottleneck, mandates a small fixed size DR i.e. a computer server servicing a small number of robots to reduce contention and transmission time. Since communication between servers is not constrained, we can have the number of servers scaling linearly with system size, giving Model 1 (completely modular network) as the optimal architecture.
\item Unlimited Robot Bandwidth, Limited Server Bandwidth: Assuming communication between computer servers is a bottleneck, stipulates a fixed number of computer servers to reduce communication overhead. Since communication between robots and servers is not constrained, we can have the DR size (number of robots serviced by a single server) and LN size (number of agents in a single server) scaling with system size. Hence the optimal architecture is Model 2 (non-modular detection network).
\item Limited Robot Bandwidth, Limited Server Bandwidth: A bottleneck in robot and server communication demands a small DR and lots of servers, whereas a bottleneck in server communication requires a large server with fewer total number of servers. The architecture that balances these opposing requirements is Model 3 (hybrid sub-modular architecture) i.e. the number of servers and their size (number of robots serviced by each server) increases with system size (\ref{fig:Fig4}).
\end{enumerate}

The local communication time within an artificial DR is a function of the number of robots ($d$) serviced by a single artificial LN 
\begin{equation}
t_{local} = f(d)
\end{equation}
 
The function $f$ will depend on constraints on communication between robots and servers, influenced, for example, by how robot requests are queued on the server and the distance over which low power robots can send and receive messages. The global communication time between artificial LNs is also a function of the number of LNs in the system ($n/d$) where $n$ is the total number of robots in the entire system

\begin{equation}
t_{global} = g(n/d)
\end{equation}

The function $g$ depends on communication constraints between servers. For low latency and high bandwidth connections among servers, $t_{global}$ may not scale appreciably. However, low power servers distributed in remote environments, may preclude broadcast communication such that $t_{global}$ increases with $n/d$. An increase in the size of an artificial LN (and hence the number of robots serviced, $d$) would reduce $t_{global}$ at the cost of $t_{local}$. The size and number of artificial LNs to balance local and global communication depends on the precise functions $f$ and $g$ mediating local and global communication.

Although we have provided only one example, this research is widely applicable to other distributed systems AIS applications. In recent work, we have extended our work to modify peer-to-peer systems with a sub-modular architecture \citep{icaris_modular}. 

In summary, understanding the tradeoff between fast search for pathogens and fast production of antibodies is important for AIS that mimic the NIS. If the goal of an AIS is only search or detection in physical space with a local response, then a completely modular design (Model 1) will be optimal. If an AIS searches in physical space but requires a global response after detection, a sub-modular architecture (Model 3) optimizes the tradeoff between local search and global response and will lead to faster search and response times ($T = t_{local} + t_{global}$). Our analysis sheds light on the relationship between physical space and architecture in resource-constrained distributed systems.

\section{Conclusion}
\label{conclusion}

Host body size constrains pathogen replication rates due to the physical characteristics of transportation networks that supply infected and normal cells with energy. Host body size also constrains NIS detection and response times by increasing the physical size of search spaces. The NIS is comprised of rare antigen-specific immune system cells that it must utilize to search for initially small numbers of pathogens localized in a large physical space. The NIS solves this classic search for a "needle in a haystack" in time that is almost invariant of the size of the organism. The decentralized nature of the lymphatic network also helps in efficient pathogen detection by acting as a small volume of tissue where DCs can efficiently present antigen to T cells. The NIS must also respond to the antigen by producing antibodies (in the case of WNV) proportional to the mass of the organism. From empirical data, that time also appears independent of body size.

We use an ODE model to show that NIS response rates are independent of host body size and pathogen replication rates decrease with body size. We examine three different hypothesized NIS architectures to explain the scale-invariant detection and response times of the NIS. The sub-modular detection network strikes a balance between the two opposing goals of antigen detection (local communication) and antibody production (global communication), and is consistent with observed numbers and sizes of LN. This is surprising since many other components of the NIS, like collections of immune system cells representing a complete repertoire, are theorized to be constructed in modular units called \textit{protectons} \citep{protecton}.

The mechanisms used by the NIS to overcome physical constraints of system size are worthy of consideration in AIS domains. Similar to NIS, physical space and resource are also important constraints on AIS, especially distributed systems applications used to connect low-powered sensors using short-range wireless communication. A sub-modular architecture efficiently balances local and global communication in AIS problems like distributed robot control that require a tradeoff between local search for a solution and global response or distribution of the solution between different components.

\section{Acknowledgements}
We would like to acknowledge fruitful discussions with Dr. Alan Perelson, Dr. Stephanie Forrest, Dr. Ruy Ribeiro, Dr. Jedidiah Crandall and Kimberly Kanigel, helpful reviews from the ICARIS 2009 conference and four anonymous referees for their insightful comments. We are also grateful to Dr. Nicholas Komar for sharing his experimental data with us. MEM and SB were supported by a grant from the National Institute of Health (NIH RR018754). SB would also like to acknowledge travel grants from RPT, SCAP and PIBBS at the University of New Mexico.

\bibliographystyle{spbasic}
\bibliography{swarm2010}   

\end{document}